\documentclass[12pt]{article}
\usepackage{graphicx}
\usepackage{cite}
\pdfoutput=1
\newcommand{\ie}{\mbox{\it i.e.}}
\newcommand{\eg}{\mbox{\it e.g.}}
\newcommand{\etal}{\mbox{\it et al.}}
\renewcommand{\epsilon}{\varepsilon}
\newcommand{\Tc}{\mbox{$T_c$}}
\newcommand{\Ds}{\mbox{$\Delta\sigma_{SF}(T,0)$}}
\newcommand{\DsH}{\mbox{$\Delta\sigma_{SF}(T,H)$}}

\newcommand{\xicdeo}{\mbox{$\xi_c(0)$}}

\newcommand{\YBCOf}{\mbox{YBa$_2$Cu$_3$O$_{6+x}$}}

\newcommand{\LSCOf}{\mbox{${\rm La}_{2-x}{\rm Sr}_x{\rm Cu}_2{\rm O}_{4}$}}
\newcommand{\CuOdosf}{\mbox{${\rm Cu}{\rm O}_{2}$}}
\newcommand{\Gsim}{\stackrel{>}{_\sim}}
\newcommand{\Lsim}{\stackrel{<}{_\sim}}
\newcommand{\eC}{\mbox{$\epsilon^C$}}

\newcommand{\onlinecite}[1]{\cite{#1}}

\begin{document}

\newcommand{\titulo}{
{Comment on}  {\it ``High-field studies of superconducting fluctuations in high-$T_c$ cuprates: Evidence for a small gap distinct from the large pseudogap''}
}

\newcommand{\autor}{M.V. Ramallo, C. Carballeira, R.I. Rey, J. Mosqueira and F. Vidal}

\newcommand{\direccion}{Laboratorio de Baixas Temperaturas e Supercondutividade LBTS,\\ Departamento de F\'{\i}sica da Materia Condensada,\\ Universidade de Santiago de Compostela, ES-15782 Santiago de Compostela, Spain}

\begin{center}\Large\bf\titulo\\ \mbox{}\\ \end{center}

\begin{center}\normalsize\autor\end{center}

\begin{center}\normalsize\it\direccion\end{center}

\mbox{}\vskip1.0cm{\bf Abstract. }
\setlength{\baselineskip}{18pt}

By using high magnetic field data to estimate the background conductivity, Rullier-Albenque and coworkers have recently published [Phys.~Rev.~B {\bf 84}, 014522 (2011)] experimental evidence that the in-plane paraconductivity in cuprates is almost independent of doping. In this Comment we also show that, in contrast with their claims, these useful data may be explained at a quantitative level in terms of the Gaussian-Ginzburg-Landau approach for layered superconductors, extended by Carballeira and coworkers to high reduced-temperatures by introducing a total-energy cutoff [Phys.~Rev.~B {\bf 63}, 144515 (2001)]. When combined, these two conclusions further suggest that the paraconductivity in cuprates is conventional, \ie, associated with fluctuating superconducting pairs above the mean-field critical temperature.

\newpage
\setlength{\baselineskip}{18pt}

In a recent work,\cite{c1} Rullier-Albenque and coworkers present detailed measurements of the in-plane paraconductivity, \Ds, and of the fluctuation-induced magnetoconductivity, \DsH, above the superconducting transition temperature, \Tc, of \YBCOf\ superconductors, as a function of the oxygen content and with magnetic fields up to 60~Tesla. As they estimate the background contributions to the resistivity by means of the high reduced-magnetic field data, instead of the usual temperature extrapolations\cite{c2}, their results provide an useful confirmation of earlier findings obtained by Curr\'as and coworkers \cite{c3} and then by other authors\cite{c4,c5}: even in the high reduced-temperature region, for  $\epsilon\equiv\ln(T/\Tc)\Gsim0.1$ (where the influence of the opening of a pseudogap in the normal state of the underdoped cuprates could be more important), \Ds\ is almost independent of the doping level. As already stressed in Refs.~\onlinecite{c3} and~\onlinecite{c6}, this conclusion, that applies to all samples not severely affected by inhomogeneities, was inferred in these different works directly from the \Ds\ data and, therefore, is ``model independent''. These results strongly suggest that the in-plane paraconductivity in cuprates is, including its onset,\cite{c7} independent of the opening of a pseudogap in the normal state.

The next crucial step to establish the nature itself of the corresponding superconducting fluctuations is to check if the measured \Ds\ could be described in terms of the different versions of the phenomenological Gaussian-Ginzburg-Landau (GGL) approach for layered superconductors. This task was also attempted in Ref.~\onlinecite{c1} but, unfortunately, without taking into account the multilayering effects\cite{c8,c9} and also, when analyzing their data on the grounds of the so-called {\it extended GGL approach} that includes a {\it total-energy cutoff}\/\cite{c10,c11}, by using an expression for \Ds\  inadequate for the studied compounds. By overcoming these shortcomings, we will show here that contrarily to the conclusions in Ref.~\onlinecite{c1} the in-plane paraconductivity data are in excellent quantitative agreement, also in the high-$\epsilon$ region, with the extended GGL approach, providing then an interesting confirmation of previous analyses obtained in other cuprates with different dopings\cite{c3,c4,c10,c11,c12}. Note also that the onset of the in-plane paraconductivity in cuprate superconductors with different dopings is still at present a central and debated aspect that is receiving considerable attention (see, \eg, Ref.~\onlinecite{c13}). So, probably one of the most useful implication of the experimental data presented in Ref.~\onlinecite{c1} is to provide, when correctly analyzed, a further confirmation of the adequacy of the GGL approach to describe the onset of the superconducting fluctuations. 

Let us note that the total-energy cutoff, introduced in the GGL calculations of the paraconductivity by Carballeira and coworkers\cite{c10}, directly results from the limits imposed by the uncertainty principle to the shrinkage of the superconducting wave function above the superconducting transition, as proposed by Vidal and coworkers\cite{c11}. This cutoff has then a fundamental origin, and it solves the well-known inconsistencies at large $\epsilon$ of the GGL approach without any cutoff or with the popular momentum cutoff, while recovering for low  $\epsilon$  the conventional (without a cutoff) GGL results, as explained in detail in Refs.~\onlinecite{c3}, \onlinecite{c10} and~\onlinecite{c11}. So, one may note already here without the need of any detailed comparison that, contrarily to the conclusion suggested in the Section~\mbox{VIII-A} of Ref.~\onlinecite{c1}, the introduction of the {\it intrinsic-like} total-energy cutoff ``privileges'' the high-$\epsilon$  region (``the cut-off behaviour'') but without appreciably affecting the low-$\epsilon$  region: In the case of the paraconductivity, the influence of such a cutoff is almost inappreciable in the low reduced-temperature region, for $\epsilon\Lsim0.03$,  it becomes relatively moderate (could be absorbed by slightly changing the parameters involved) for $0.03\Lsim\epsilon\Lsim0.1$, and it only becomes important in the high reduced-temperature region, for $\epsilon\Gsim0.1$. 

A comparison between the data of Fig.~24 and Eq.~(24) in Ref.~\onlinecite{c1} would already provide a crude confirmation of the qualitative considerations indicated above, even when using as effective interlayer periodicity length, $s$, the crystallographic unit cell length, $c=11.7$\AA\ for \YBCOf. However, Eq.~(24) agrees only in the 2D-limit with the expression calculated by Carballeira and coworkers for the so-called {\it direct} (Aslamazov-Larkin) contribution to \Ds\ under a total-energy cutoff\cite{c3,c10}. This limit is defined by $2\xicdeo\epsilon^{-1/2}\ll s$, where \xicdeo\ is the transversal coherence length amplitude. In \YBCOf\ such a limit will not apply when $\epsilon\Lsim0.1$. So, to perform a quantitative comparison, we must use the general expression for \Ds\ given by Eq.~(9) of Ref.~\onlinecite{c10}, and also one must take into consideration that these compounds have two superconducting layers per unit cell length and that the corresponding multilayering effects may affect both the amplitude and the $\epsilon$-dependence of the in-plane paraconductivity\cite{c8,c9}. As it is now well established\cite{c9,c15}, in the case of \YBCOf\ these multilayering effects may be crudely taken into account by just using an effective interlayer distance $s\simeq c/2 = 5.85$\AA. This last approximation, that was already used by Carballeira and coworkers when analyzing the paraconductivity measured in optimally doped \YBCOf\ single crystals and thin films,\cite{c10} means that in these compounds the two superconducting layers in the periodicity length $c$ may fluctuate as different degrees of freedom, a conclusion that was confirmed experimentally in other works\cite{c9,c15}. Note that multilayering does not imply any value for \xicdeo\ [in particular, it does not impose at all $\xicdeo=0$] and, obviously, it does not exclude the presence of a 2D-3D transition of the superconducting fluctuations when approaching \Tc\ from above. This point was clearly stressed in Refs.~\onlinecite{c9}, \onlinecite{c10} and~\onlinecite{c15}, where by analyzing consistently three different observables it was also established experimentally for the first time the absence of indirect (Maki-Thompson) and DOS effects on the paraconductivity in cuprate superconductors.

The comparison between the experimental data of Fig.~24 in Ref.~\onlinecite{c1} and Eq.~(9) in Ref.~\onlinecite{c10} is presented in the Fig.~1 of this Comment. We have used $s = c/2$, and for the total-energy cutoff parameter\cite{c10,c11} $\eC=0.5$, which is close to the value estimated in the BCS scenario\cite{c11}. For the remaining parameter, \xicdeo, we have used the same value as proposed in Ref.~\onlinecite{c1}, $\xicdeo=0.9$\AA\ (which is close to the value used in Refs.~\onlinecite{c9} to~\onlinecite{c11} for optimally doped \YBCOf). The resulting \Ds\ is the upper solid curve in Fig.~1. As one may appreciate in that figure, the in-plane paraconductivity data for the different \YBCOf\ superconductors presented in Fig.~24 of Ref.~\onlinecite{c1} agree at a quantitative level, in the whole range of $\epsilon$ values and doping levels covered by these data, with the GGL predictions for the direct (Aslamazov-Larkin) fluctuation contribution in bilayered superconductors under a total-energy cutoff proposed by Carballeira and coworkers in Ref.~\onlinecite{c10}. For completeness, we also present in that figure (lower solid curve) the corresponding predictions in absence of multilayering, \ie, those obtained by using in Eq.~(9) of Ref.~\onlinecite{c10} $s = c = 11.7$\AA\  and the same values as before for the remaining parameters, \ie, $\eC=0.5$ and $\xicdeo=0.9$\AA. Although for $\epsilon<0.1$ the agreement is somewhat worse than when the multilayering effects are taken into account, even in this case the improvement obtained by the introduction of the total-energy cutoff is evident, mainly when compared with the GGL predictions without cutoff (dashed curves in Fig.~1, labeled as Eq.~LD following the notation in Ref.~\onlinecite{c1}). The latter were evaluated by using Eq.~(10) of Ref.~\onlinecite{c1} with either $s=c/2$ or $s=c$.

In conclusion, by analyzing, just as an example, the results of Fig.~24 of Ref.~\onlinecite{c1}, we have shown that contrarily to the claims of Rullier-Albenque and coworkers their paraconductivity results may be explained at a quantitative level in terms of the {\it extended} GGL approach. This result further confirms, at least for dopings above 0.1~hole/\CuOdosf,\cite{c16} the conventional nature of the superconducting fluctuations in cuprates (GGL-like, associated with fluctuating superconducting pairs above the mean-field critical temperature), independently of their doping and of the temperature region above \Tc, as also earlier concluded by Curr\'as and coworkers\cite{c3} by analyzing the in-plane excess conductivity in other cuprates (see also Refs.~\onlinecite{c4} to~\onlinecite{c6}  and~\onlinecite{c10} to~\onlinecite{c12}). They also seem to confirm that the total-energy cutoff parameter is, well within the experimental uncertainties, doping-independent and close to the value that may be estimated in the BCS scenario\cite{c11}. The example studied here also suggests the way to analyze in terms of the extended GGL approach the remaining measurements of Ref.~\onlinecite{c1}. It would be also interesting to compare the data acquired under high reduced-magnetic fields with results for the fluctuation-induced diamagnetism in  \LSCOf\  superconductors with different dopings, which follow the GGL predictions even for finite reduced-magnetic fields\cite{c17}.

{\it Acknowledgments.--- }  This work has been supported by the MICINN project FIS2010-19807 and by the Xunta de Galicia projects 2010/XA043 and 10TMT206012PR. All these projects are co-funded by ERDF from the European Union.

%%% Figure 1%%%%%

\begin{figure}[ht!]
\mbox{}\vspace{0.3cm}\mbox{}\\\vspace{1cm}
\mbox{}\hfill\includegraphics[width=0.65\textwidth]{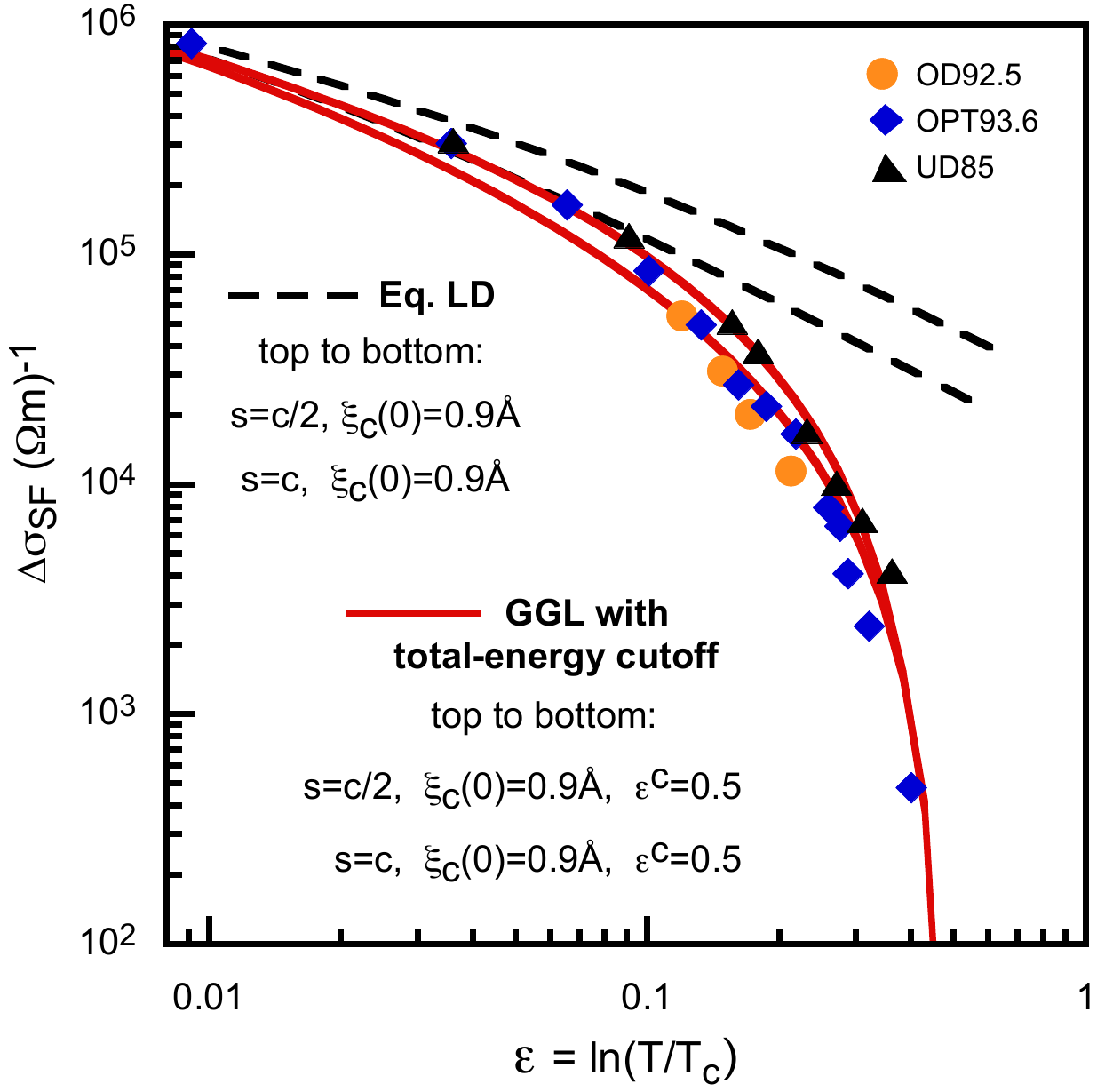}\hfill\mbox{}
\caption{\setlength{\baselineskip}{18pt}
\label{figura}
(color online) Comparison between some of the experimental data for the in-plane paraconductivity in \YBCOf\ superconductors with different dopings, summarized in Fig.~24 of Ref.~\onlinecite{c1}, with the expression for the GGL paraconductivity under a total-energy cutoff, calculated by Carballeira and coworkers for multilayered superconductors [Eq.~(9) of Ref.~\onlinecite{c10}]. In doing this comparison we have used the $c$-direction coherence length proposed in Ref.~\onlinecite{c1}, an effective interlayer distance $s = c/2=5.85$\AA, and a total-energy cutoff parameter $\eC=0.5$, which is close to the value that one may estimate in the BCS scenario\cite{c11}.  We also show (upper dashed line) the GGL prediction without cutoff [Eq.~(10) of Ref.~\onlinecite{c1}], evaluated by using again $s = c/2$, and that following the notation used in Ref.~\onlinecite{c1} is labeled Eq.~LD in this figure. The lower continuous and dashed curves are the in-plane paraconductivity neglecting multilayering effects, \ie, calculated by using $s=c$ and the same values as before for the remaining parameters. }
\end{figure}


\begin{thebibliography}{99}
\bibitem{c2}{
The need of an always somewhat arbitrary background term to determine the paraconductivity is a consequence of the fact that the theoretical approaches calculate the fluctuation contribution separately from the non-critical contributions. These ambiguities affect, therefore, any background estimation, including those obtained by quenching the superconductivity by high reduced-magnetic fields. So, the comparison between the paraconductivities obtained by using different background procedures provides then an important test of consistency.
}
\bibitem{c7}{
The paraconductivity behavior at high reduced-temperatures, including the temperature location of its onset, is particularly sensitive to the background estimations, and the use of linear temperature extrapolations may introduce considerable errors. This point was clearly stressed in Ref.~\onlinecite{c3}, where it was detailed a procedure to avoid these shortcomings when the normal-state resistivity is nonlinear. Although these difficulties have been described in detail in different works (see, \eg, Refs.~\onlinecite{c3} to~\onlinecite{c6} and references therein), some of them were revisited in Ref.~\onlinecite{c1} (in Section VI). In any case, the central aspect here is the fact that these different background procedures, including the one of Ref.~\onlinecite{c1}, lead to a doping-independent behavior of the in-plane paraconductivity at high reduced-temperatures, as first observed in Ref.~\onlinecite{c3}. In view of this conclusion, any debate about the best procedure to estimate the background becomes irrelevant here.
}
\bibitem{c14}{
In contrast with the claims of Rullier-Albenque and coworkers, their Eq.~(24) equals the result of Ref.~\onlinecite{c10} for layered superconductors only in the 2D limit. However, one cannot attribute to this mistake the erroneous qualitative conclusions of Section~\mbox{VIII-A} of Ref.~\onlinecite{c1}, as for these compounds the resulting predictions will differ only moderately. Nevertheless, for quantitative analyses one must use the general expression for the direct (Aslamazov-Larkin) fluctuation contribution to the in-plane paraconductivity, given by Eq.~(9) in Ref.~\onlinecite{c10}. Note that the proposals in such Ref.~\onlinecite{c10} are based on phenomenological but fundamental grounds [instead of data fittings, as it seems to be the case of Eq.~(25) in Ref.~\onlinecite{c1}]. 
}
\end{thebibliography}
\end{document}